\documentclass[journal]{IEEEtran}
\usepackage{units}
\usepackage{ifpdf}
\usepackage{cite}
\ifCLASSINFOpdf
  \usepackage[pdftex]{graphicx}
\else
  \usepackage[dvips]{graphicx}
\fi
\usepackage[cmex10]{amsmath}
\usepackage{array}
\usepackage{mdwtab}
\usepackage[font=footnotesize,caption=false]{subfig}
\usepackage{url}
\begin{document}

\title{Connectivity statistics of store-and-forward inter-vehicle communication}

\author{Arne~Kesting, Martin~Treiber,\thanks{A. Kesting and M. Martin are with the Department of Transport and Traffic Sciences, Technische Universit\"at Dresden, W\"urzburger Str. 35, 01187 Dresden, Germany. e-mail: (see http://www.akesting.de and http://www.mtreiber.de).}
and~Dirk~Helbing\thanks{D. Helbing is with ETH Zurich, UNO D11,
CH-8092 Zurich, Switzerland and Collegium Budapest -- Institute for
Advanced Study, Szenth\'aroms\'ag u. 2, H-1014 Budapest,
Hungary. e-mail: (see http://www.helbing.org).}}

\markboth{IEEE Transactions on Intelligent Transportation Systems,~Vol.~11, No.~1, 172-181, 2010}%
{Kesting \MakeLowercase{\textit{et al.}}: Connectivity statistics of store-and-forward inter-vehicle communication}

\maketitle

\begin{abstract}
Inter-vehicle communication (IVC) enables vehicles to exchange messages within a limited broadcast range and thus self-organize into dynamical vehicular ad hoc networks. For the foreseeable future, however, a direct connectivity between equipped vehicles in one direction is rarely possible. We therefore investigate an alternative mode in which messages are stored by relay vehicles traveling in the opposite direction, and  forwarded to vehicles in the original direction at a later time. The wireless communication consists of two ``transversal'' message hops across driving directions. Since direct connectivity for transversal hops and a successful message transmission to vehicles in the destination region is only a matter of time, the quality of this IVC strategy can be described in terms of the distribution function for the total transmission time.  Assuming a Poissonian distance distribution between equipped vehicles, we derive analytical probability distributions for message transmission times and related propagation speeds for a deterministic and a stochastic model of the maximum range of direct communication. By means of integrated microscopic simulations of communication and bi-directional traffic flows, we validated the theoretical expectation for multi-lane roadways. We found little deviation of the analytical result for multi-lane scenarios, but significant deviations for a single-lane. This can be explained by vehicle platooning.  We demonstrate the efficiency of the transverse hopping mechanism for a congestion-warning application in a microscopic traffic simulation scenario. Messages are created on an event-driven basis by equipped vehicles entering and leaving a traffic jam. This application is operative for penetration levels as low as 1\%.
\end{abstract}

\begin{IEEEkeywords}
Inter-vehicle communication, connectivity, vehicular ad-hoc networks, traffic simulation 
\end{IEEEkeywords}

\IEEEpeerreviewmaketitle

\section{\label{sec:intro}Introduction}
\IEEEPARstart{I}{nter-vehicle} communication (IVC) based on wireless communication
among vehicles is widely regarded as a promising application for novel
transportation services with applications in traffic safety, advanced
traveler information and driver assistance
systems~\cite{Shladover2007-TRR-IVC,wisch,zhang2005-IVC,Mahmassani2001-DNA,Arne-ACC-TRC,Kato2002-IVC,Arem_ACC_Impact_Transaction}. In
contrast to conventional communication channels which operate with a
centralized broadcasting concept via radio or mobile-phone services,
vehicular ad hoc networks (VANETs) provide a {\it decentralized
approach} that does not rely on public infrastructure. Since direct
wireless communication has a limited range of, typically, a few
hundreds of meters, the effectiveness of potential applications
depends crucially on the market penetration. The percentage of
equipped vehicles is expected to remain well below~10\% for the next
years, so any realistic concept is required to operate at penetration
levels of a few percent.

Wireless local area networks based on IEEE~802.11 have shown broadcast
ranges between~200 and~\unit[500]{m} in automobile
applications~\cite{Singh2002,OttKut2004}. However, VANETs have several
characteristics that distinguish them from other ad hoc
networks. Among those is the potential change in the node distribution
due to the movement of sender/receiver vehicles. For this reason,
dedicated protocols and transmission standards are intensively
investigated~\cite{korkmaz2004umh,artimy2007-Transactions,saito2007-IVC,Sen_IVC_5GHz}. Detailed
network simulation tools have also been applied to VANET in order to
examine the influences of environmental conditions and node mobility
on connectivity~\cite{torrent-moreno2004brr,Moreno-IVC-Failure}. For
an overview of the technical architectures and protocols, and current
research projects and consortia initiatives, we refer to the survey of
Hartenstein and Laberteaux~\cite{Hartenstein-IVC-Survey}.

In most automobile applications, it is necessary to carry messages
over distances that are significantly longer than the device's
broadcast range meaning that, in general, several equipped vehicles
acting as relays are necessary to transport the message to the final
destinations. Depending on the application and the specific IVC
strategy, a certain minimum percentage of equipped cars is necessary
for operation. This characterizes the fundamental problem of a
\textit{market penetration threshold}. There are two basic strategies
of message propagation:
\begin{enumerate}
 \item[(1)] A message can be passed backwards to the following
 equipped vehicle, which then passes it to the next equipped vehicle,
 and so on, resulting in a node connectivity via multiple hops.  We
 call this \textit{longitudinal hopping}, as the message always
 propagates parallel to the travel direction of the first sender and
 the last receiver.  

 \item[(2)] A sender may transfer a message to a vehicle driving in
 the opposite direction. This vehicle can store the message and
 continuously broadcast it for a certain period of time, while
 physically transporting the message upstream.  Although this message
 is of no use for the relay vehicle, it might eventually be received
 by an equipped vehicle driving in the original direction by a second
 transversal hop. We therefore call this strategy ``store and
 forward'', or \textit{transversal hopping}.
\end{enumerate}
The longitudinal hopping mode has the advantage of virtually
instantaneous message transmission, so the transmitted information is
always up-to-date. However, the longitudinal communication chain
typically fails when the broadcast range~$r$ is of the same order or
smaller than the typical distance between equipped vehicles. For
realistic broadcast ranges of a few hundreds meters the penetration
threshold is about 10\% for typical densities on
freeways~\cite{thiemann-IVC-PRE08}. This restriction can be overcome
by the transversal hopping mode, for which the successful transmission
is only a matter of time, but the information may be already obsolete
when it finally arrives. 
The direct connectivity is determined mainly by the broadcast range, 
the traffic density, the market penetration of
IVC vehicles and the distribution of equipped vehicles in the traffic
stream. The traffic density characterizes the global number of
vehicles, but varies strongly in congested traffic conditions. We will
therefore focus on the influence of the traffic dynamics on the
connectivity of the ad-hoc network of equipped cars. This delay-based connectivity ist most useful for 
traveler traffic information services rather than safety applications.

Previous work on connectivity has been mainly focused on instantaneous connectivity. In the literature, the  multiple hopping of messages in longitudinal direction has been studied by several
authors~\cite{Jin-IVC-2006,Jin2007-IVC,wang2007-IVC-TRB,Ukkusuri-IVC-TRC,thiemann-IVC-PRE08}.
Due to the inherent complexity of mobile ad hoc communication within
the traffic stream, realistic simulations are also essential for
validating the analytical models and testing IVC use
cases~\cite{Lovell-IVC-2006,choffnes2005ima,thiemann-IVC-PRE08,artimy2007-Transactions,Fiore-VANET-SIM-2007}. Fewer
investigations considered the transverse propagation process by means
of simulation~\cite{Yang2005,IVC-Martin-TRR07}.  To our knowledge, no
analytical models have been proposed or investigated for this type of
IVC message transport.

In this paper, we will consider the transverse message propagation
mode in which messages are transported by vehicles traveling in
opposite direction. The system and the relevant transmission processes
are described in Sec.~\ref{sec:trans}. In Sec.~\ref{sec:model_trans},
we will derive analytical probability distributions for message
transmission times and related propagation speeds, assuming a
Poissonian distance distribution between equipped vehicles. By means
of microscopic traffic simulations combining communication and
bi-directional traffic flows, we test the theoretical predictions with
numerical results for single and multi-lane roadways in
Sec.~\ref{sec:sim_trans}. In the subsequent section, we simulate an
incident scenario and consider the propagation of information on
shock-fronts and travel times to vehicles in the upstream direction,
and discuss future applications in traveler information and
``traffic-adaptive'' cruise control systems.   In Sec.~\ref{sec:diss},
we conclude with a discussion of our findings.

\section{\label{sec:trans}Message transport via the opposite driving direction}

In this section, we consider a concept of inter-vehicle communication,
using equipped vehicles in the opposite driving direction as
relays. These vehicles transport the message to a destination region
where it is delivered back to vehicles in the original driving
directions. Thus, the wireless communication part of the mechanism
considered consists of exactly two ``transverse message hops''
across driving directions.

In the following, a message is considered to be useful if it reaches
an equipped vehicle driving in the same direction as the vehicle
creating the initial message before the receiver reaches the position
of the original message. For example, if the initial message contains
information about a locally detected traffic jam, this information is
useful only for vehicles sufficiently upstream of this location and
driving in the same direction. Specifically, we define as
\textit{destination region} all locations in this direction that are
upstream of the original source by at least the distance
$r_\text{min}$ (typical values for $r_\text{min}$ are of the order of
\unit[1]{km}). Message transmission by the transverse hopping strategy can be
considered as a three-stage process:
\begin{enumerate}
\item[(1)]
After a message is generated by an equipped vehicle, the message is
continually broadcasted by this vehicle for a certain time. It is
received by a relay vehicle in the opposite driving direction after a
time interval $\tau_1$.
\item[(2)]
The message is then transported by the relay vehicle. After a time
interval $\tau_2$  (with respect to the initial generation of the
message) this vehicle's broadcast range starts intersecting the
destination region and the vehicle starts broadcasting the stored message.
\item[(3)]
At time $\tau_3$ (with respect to the initial generation time) the
message is received by an equipped vehicle in the destination region. The
relay vehicle continues broadcasting until the message is considered
to be obsolete (the criteria for this condition will not be considered
here).
\end{enumerate}
Because of the complex traffic dynamics, the time intervals $\tau_i$
(with $i=1, 2, 3$) can be considered as \textit{stochastic variables}.

\begin{figure}[!t]
 \centering
 \includegraphics[width=\linewidth]{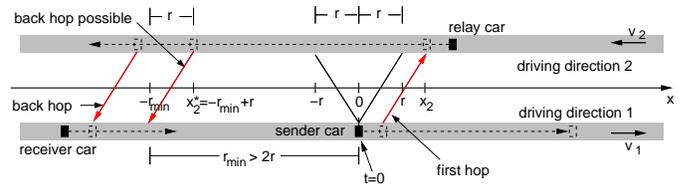}
 
  \caption{\label{fig:trans_sketch}Illustration of message transport
  via the opposite driving direction. At time $t=0$, a car in
  direction~1 generates a message and starts broadcasting it. The
  message is transported forward at speed $v_1$ until it will be
  received by an equipped car moving at speed $v_2$ in
  direction~2. The relay car transports the message at least up to the
  position $-r_\text{min}+r$ (the destination region) before broadcasting
  it again. Finally, the message is received by an equipped car in
  direction~1. Transversal distances are neglected when determining
  the communication range $r$.}

\end{figure}

\section{\label{sec:model_trans}Analytical model}
For the description of the information propagation via the opposite
driving direction, we consider a bi-directional freeway or arterial
road (cf.\ Fig.~\ref{fig:trans_sketch}).  In both directions, we
assume essentially homogeneous traffic flows that are characterized by
the lane-averaged velocities $v_1$ and $v_2$, and the total densities
$\rho_1$ and $\rho_2$, respectively.  Furthermore, we assume a global
market penetration $\alpha$ such that the relevant density variables
are the {\it partial densities of equipped vehicles}, $\lambda_1 =
\alpha
\rho_1$ and $\lambda_2=\alpha \rho_2$, respectively. Communication
between equipped vehicles is possible within a limited broadcast range
$r$. Within this range, the communication is assumed to be error free
and instantaneous. In Sec.~\ref{sec:stoch}, we will relax this
assumption and derive an analytical model for the more realistic case
of distributed broadcast
ranges~\cite{Moreno-IVC-Failure,Hartenstein-IVC-Survey}.  By virtue of
the assumed constant velocities, the first relay vehicle having
received the message will also be the first vehicle reaching the
destination region, so we will ignore message reception and transport
by subsequent relay vehicles.

Let us consider a message that is generated by an equipped vehicle
driving in direction~1 at position $x=0$ and time $t=0$. As
illustrated in Fig.~\ref{fig:trans_sketch}, we use the same coordinate
system for both directions such that driving direction~1 is parallel
and direction~2 anti-parallel to the $x$-axis.  We start with the
spatial distribution of vehicles in the opposite driving direction.
We assume that the distances between {\it equipped} vehicles are
i.i.d.\ exponentially distributed stochastic variables which is a good
approximation as long as the partial densities $\lambda_1$ and
$\lambda_2$ are small~\cite{thiemann-IVC-PRE08}.
Because of the implied Poisson process, the same distribution applies
to the distance between \textit{any} given location $x$ and the next equipped vehicle.

\subsection{First Transversal Hop}
First, we investigate the statistical properties of the time $\tau_1$
of the first transversal hop to a relay vehicle in the opposite
direction~2. Since only the \textit{first} transmission of a message
is relevant, the reference point for finding such a vehicle is given
by the ``best case''. In this case, the initial position of the relay
is at $x_2=-r$, i.e., at the limit of the transmission range in the
desired direction.  The actually used transmitter vehicle is the
equipped vehicle located next to this point when going in the
direction of positive $x$. According to the above
assumptions, its initial position $X_2$ is a stochastic variable that
is defined by the probability density
\begin{equation}\label{eq:poisson_distr}
f(x) = \lambda_2 e^{-\lambda_2(x+r)}\Theta(x+r).
\end{equation}
Here, the Heaviside function $\Theta(x)$ (which is equal to~$0$ for $x<0$ and equal to~1
for $x\ge 0$) serves as a cut-off for receiver positions at $x<-r$,
which are too remote. Notice that a ``conventional''
exponential distribution would result when starting the search for a
relay vehicle at $x=0$, i.e., setting $r=0$.

An instantaneous transversal hop of the generated message to the relay
vehicle on roadway~2 is possible if it is within the broadcast range,
$-r \le X_2 \le r$. This happens with probability
\begin{equation}
P_1(0) = \int_{-r}^{r} f(x)\, \text{d}x = 1-e^{-2\lambda_2 r}.
\end{equation}
Otherwise, a finite time interval $\tau_1$ is needed before the relay
vehicle comes within communication range (cf.\
Fig.~\ref{fig:trans_sketch}). Since the relative velocity between
vehicles in opposite driving directions is $v_1+v_2$, this time
depends on the initial position $X_2$ according to
$\tau_1=(X_2-r)/(v_1+v_2)$. Therefore, a successful first hop is
possible at time $\tau$ or before, if the initial position of the
receiving vehicle is in the interval $X_2 \in [-r, x_2(\tau)]$ where
$x_\text{2}(\tau) = (v_1+v_2) \tau + r$. Consequently, the
distribution function for $\tau_1$ (the probability for a successful
first hop at time $\tau_1$ or earlier) is given by
\begin{equation}\label{eq:prob_1}
P_1(\tau) =\int_{-r}^{x_\text{2}(\tau)} f(x)\, \text{d}x =
1-e^{-\lambda_2\left[2 r + (v_1+v_2) \tau \right]}.
\end{equation}
This probability is shown in Fig.~\ref{fig:trans_statistics}(a) for
different market penetrations $\alpha$ influencing the partial density
of equipped cars, $\lambda_2=\alpha \rho_2$.

\subsection{Possibility of Hopping to Original Driving Direction}
\begin{figure*}[!t]
\centering
\subfloat{\includegraphics[width=2.2in]{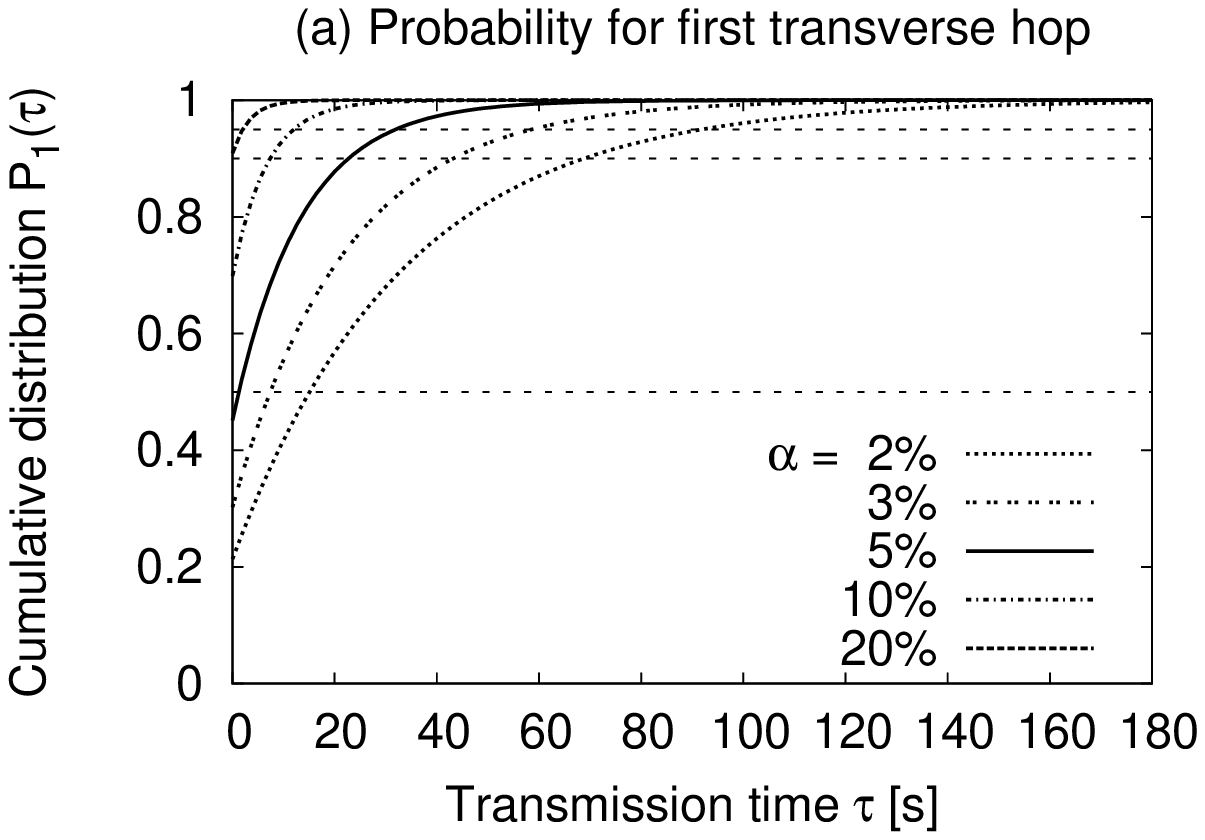}}
\subfloat{\includegraphics[width=2.2in]{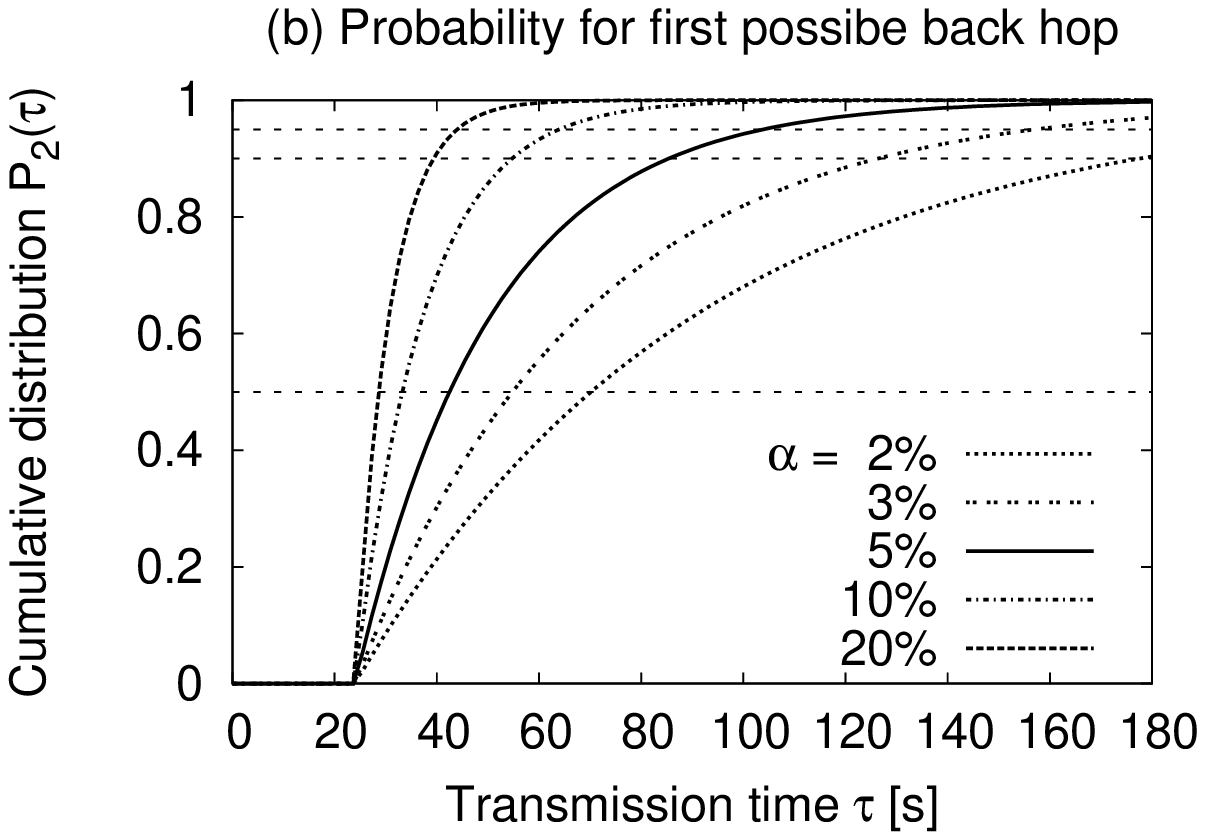}}
\subfloat{\includegraphics[width=2.2in]{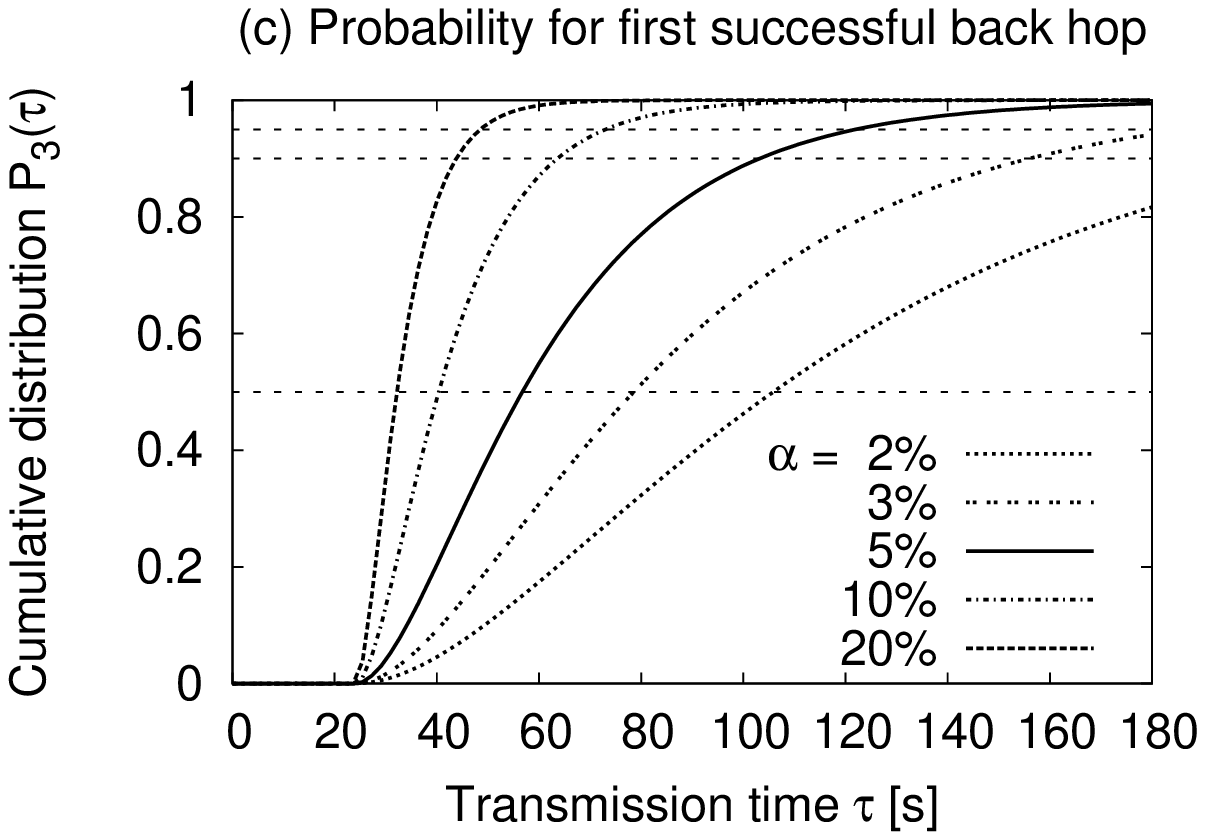}}

  \caption{\label{fig:trans_statistics}Statistical properties of the
  characteristic times for the transverse hopping mechanism. The plots
  show (a) the distribution functions for the time interval $\tau_1$
  between message creation and the first transverse hop, (b) the time
  interval $\tau_2$ after which the message becomes available in the
  destination region, and (c) the time $\tau_3$ of the first reception by a
  car in the destination region. The market penetration level $\alpha$ is
  varied, while the other parameters are kept constant (broadcast range
  $r=\unit[200]{m}$, traffic densities $\rho_1=\rho_2=\unit[30/]{km}$,
  average speeds $v_1=v_2=\unit[90]{km/h}$).}

\end{figure*}

Now we investigate the distribution of the time interval $\tau_2$
after which the message is available for retransmission to the
original driving direction~1, i.e., the relay vehicle starts
broadcasting the message. It should occur at a location $x_3$ that,
for this direction, is sufficiently upstream of the original message
source, i.e., $x_3 \le - r_\text{min}$. The transmission becomes
possible when the relay vehicle passes the position
$x_2^*=-r_\text{min}+r$. Because of the assumed constant velocities
$v_2$ in negative $x$-direction, the position of the relay vehicle
depends on time according to $X_2(\tau)=X_2-v_2 \tau$, where the
initial position $X_2$ is distributed according to the
density~\eqref{eq:poisson_distr}.  Notice that, by virtue of the
constant velocity $v_2$, the distribution function of the position of
the relay vehicle is shifted uniformly in time. Starting directly from
its definition, the distribution function for $\tau_2$ is given by
$P_2(\tau)=P(X_2(\tau)<x_2^*)=P(X_2<v_2\tau+x_2^*)=F(v_2\tau+x_2^*)$,
where $F(x)$ is the cumulative distribution function of the
probability density~\eqref{eq:poisson_distr}. Inserting this equation
results in
\begin{multline}\label{eq:prob_2}
P_2(\tau) = \int\limits_{-r}^{v_\text{2}\tau +x_2^* } f (x) \,\text{d}x = \\
 \Theta\left(\tau - \frac{r_\text{min}-2 r}{v_2} \right)
\left(1-e^{-\lambda_2(2 r + v_2\tau -r_\text{min})} \right).
\end{multline}
Since the message dissemination depends on the relay car in
direction~2 only, the probability~\eqref{eq:prob_2} is independent of
the average vehicle speed $v_1$ in direction~1 and the time $\tau_1$
of the first hop. Figure~\ref{fig:trans_statistics}(b) shows the
cumulative distribution $P_2(\tau)$ for different market penetrations
$\alpha$.

\subsection{Successful Message Transmission to Original Driving Direction}
Finally, we calculate the distribution of the time $\tau_3$ for the
first successful message receipt by a vehicle traveling in the
destination region, i.e., in direction~1 at a position $x\le
-r_\text{min}$. In general, this time differs from $\tau_2$ since, at
this time, the message can be received only by a vehicle located
exactly at $x=-r_\text{min}$. For $t>\tau_2$, the average arrival rate
of equipped vehicles in the destination region, reaching the range of the
available message, is given by the relative flow $\lambda_1(v_1+v_2)$.
For the assumed Poissonian process, the time interval
$T=\tau_3-\tau_2$ for the arrival of the first receiver vehicle is
exponentially distributed with the probability density
\begin{equation}
\label{fT}
f_T(\tau)=\Theta(\tau)\, \lambda_1 (v_1+v_2) \, e^{-\lambda_1 (v_1+v_2) \tau}.
\end{equation}
Furthermore, by means of the same Poissonian assumption, the
stochastic variables $T$ and $\tau_2$ are independent, and the
distribution function $P_3(\tau)=P(\tau_3<\tau)$ for the sum
$\tau_3=\tau_2+T$ is, therefore, given by the convolution
\begin{equation}
\label{eq:convol}
P_3(\tau)=\int\limits_0^{\infty} f_T(t) P_2(\tau-t)\, \text{d}t.
\end{equation}
Here, the Heaviside function of Eq.~\eqref{fT} has been used to limit
the lower boundary of the integral. Inserting the density function
$f_T$ and expression~\eqref{eq:prob_2} results in
\begin{multline}
P_3(\tau) =\Theta(\tau-\tau_\text{min}) \, 
\lambda_1(v_1+v_2) \int\limits_0^{\tau-\tau_\text{min}} 
e^{-\lambda_1 (v_1+v_2) t } \\ 
\left(1-e^{-\lambda_2 v_2(\tau-\tau_\text{min}-t)}\right)\,\text{d}t
\end{multline}
where the minimum transport time for a successful complete
transmission is given by $\tau_\text{min}=(r_\text{min}-2r)/v_2$.
Notice that the function $\Theta(\tau-\tau_\text{min})$ ensures that
the upper integral limit is larger than the lower one. Finally, basic
integration and introduction of the abbreviation
$\tilde{\lambda}_1=\lambda_1\frac{v_1+v_2}{v_2}$ yield the solution
\begin{multline}\label{eq:prob_3}
P_3(\tau) = \Theta(\tau-\tau_\text{min}) \\
\left( 1- \frac{\tilde{\lambda}_1}{\tilde{\lambda}_1 - \lambda_2} e^{-
\lambda_2 x_e(\tau)} + \frac{\lambda_2}{\tilde{\lambda}_1 - \lambda_2}
e^{-\tilde{\lambda}_1 x_e(\tau)}\right),
\end{multline}
where $x_e(\tau)=v_2(\tau-\tau_\text{min}) =v_2\tau - r_\text{min}+2r$
denotes the part of the destination region that intersects (or has been
intersected by) the range of the relay vehicle for the ``best case''.
In case of identical traffic conditions in both driving directions
(i.e., $v_1=v_2=v$ and $\lambda_1=\lambda_2=\lambda$), we have
$\tilde{\lambda}_1=2\lambda$, resulting in the more intuitive
expression
\begin{equation}\label{eq:prob_3_simple}
P_3(\tau) = \Theta\left(\tau - \frac{r_\text{min}-2 r}{v} \right)
\left[ 1-e^{-\lambda(2 r + v \tau -r_\text{min})}\right]^{2}
\end{equation}
which is shown in Fig.~\ref{fig:trans_statistics}(c). Note that the
quadratic term in Eq.~\eqref{eq:prob_3_simple} reflects the fact that
two encounters of equipped vehicles are needed for propagating a
message by means of transverse message hopping.

For an evaluation and discussion of the obtained results, we consider
characteristic quantities of the obtained probability distributions
which, for simplicity, are given for the symmetric case reflected by
Eq.~\eqref{eq:prob_3_simple}. A suitable measure for assessing the
performance of the proposed communication scheme are the quantiles
$\tau_3^{(q)}$, indicating the total communication time that is only
exceeded by the fraction $1-q$ of all message transports (cf.\ the
horizontal lines in Fig.~\ref{fig:trans_statistics}). The defining
condition $P_3(\tau^{(q)})=q$ leads to
\begin{equation}
\tau_3^{(q)} =  \frac{r_\text{min}-2 r}{v} 
+ \frac{ \ln\left(1-\sqrt{q} \right)}{\lambda v}. 
\end{equation}
Table~\ref{tab:trans_analyt} lists the values for the median
($q=0.5$), the 90\% and the 95\% quantiles, respectively, using the
parameters considered in Fig.~\ref{fig:trans_statistics}. For example,
even for a penetration level as low as 2\%, half of all message
transports have been completed (which is defined by the fact that a
car in the destination region at least 1~km upstream of the information
source has received the message) in~\unit[106]{s} or less. Moreover,
90\% of the propagated messages have been completed in~\unit[222]{s}
or less and only for~5\% of all messages the communication time exceeds
about~\unit[5]{min}, after which one could consider the information as
obsolete.

Furthermore, the Table~\ref{tab:trans_analyt} lists the expected
averages, the analytical values of which are given for the symmetric
case by
\begin{equation}\label{eq:average_msg_speeds}
 \langle \tau_2 \rangle = \frac{r_\text{min}-2 r}{v} +\frac{1}{\lambda v}, 
\quad
 \langle \tau_3 \rangle = \frac{r_\text{min}-2 r}{v} +\frac{3}{2 \lambda v}.
\end{equation}
The latter quantity leads us to the definition of the average speed of
information dissemination:
\begin{equation}
v_p = \frac{r_\text{min}}{\langle \tau_3 \rangle}.
\end{equation}
We will discuss the relevance of this quantity for the spread of
information in traffic flows in Sec.~\ref{sec:appl_trans}
below.  

\subsection{\label{sec:stoch}Distributed communication ranges}
In the previous section, we have derived the probability $P_3(\tau)$ for
a successful communication after the time interval $\tau$
 assuming a fixed communication range with 100\% connectivity for
distances less than $r$, and no communication for larger distances.
In order to assess the errors made by this somewhat unrealistic
assumption, we
will now derive  the probability $P_3(\tau)$ for
distributed direct communication ranges given by the density function
$g(r)$. Assuming a perfect correlation of the ranges for the first and
the second hop (which arguably is the ``worst case'' in terms of 
deviations to the simpler model~\eqref{eq:prob_3}), the probability
$P_3^\text{dist}$ for distributed communication ranges is the weighted average
of the probability $P_3(\tau|r)$ for a given direct communication
range $r$ 
weighted with the probability density:
\begin{equation}
\label{eq:PdistGen}
P_3^\text{dist}(\tau)=\int_0^{\infty} dr\ g(r) P_3(\tau | r).
\end{equation}
This expression can be analytically integrated, e.g., for uniformly,
Gaussian, or exponentially distributed ranges. For the exponential
distribution, $g(r)=\lambda_r e^{-\lambda_r r}$ for $r \ge 0$, we obtain
\begin{multline}
\label{eq:prob_3dist}
P_3^\text{dist}(\tau) =e^{-\lambda_r r_0} \, \\
\left[ 1
-\frac{\tilde{C}_1 \lambda_r}{\lambda_r+2 \lambda_2} e^{-2 \lambda_2 r_0}
+\frac{\tilde{C}_2 \lambda_r}{\lambda_r+2 \tilde{\lambda}_1} e^{-2 \tilde{\lambda}_1 r_0}
\right]
\end{multline}
where
\begin{eqnarray}
\tilde{C}_1 &=& \frac{\tilde{\lambda}_1}{\tilde{\lambda}_1-\lambda_2}
e^{\lambda_2 (r_\text{min}-v_2\tau)},\\
\tilde{C}_2 &=& \frac{\lambda_2}{\tilde{\lambda}_1-\lambda_2}
e^{\tilde{\lambda}_1 (r_\text{min}-v_2\tau)},\\
r_0 &=& \max \left[ 0, \frac{1}{2}(r_\text{min}-v_2\tau)\right].
\end{eqnarray}

\subsection{Discussion of Model Assumptions}
Let us finally discuss the implications of the three main model
assumptions: (i) exponential headway distributions, (ii) homogeneous
traffic flow, and (iii) fixed vs. distributed communication ranges.
When relaxing the assumption of an exponential distribution
(which is not valid for high traffic densities and, simultaneously, high
market penetration levels), the lines of
the model's derivation remains valid as long as the gaps between
vehicles remain uncorrelated. However, it is to be expected that the resulting integrals are
more cumbersome to solve or cannot be analytically solved at
all. Nevertheless, we will show by means of traffic
simulations in Sec.~\ref{sec:sim_trans} that the results are
remarkably robust with respect to violations of the Poisson
assumption.  

The assumption of homogeneous traffic flow implies a more serious
restriction as, after all, traveler information about traffic
congestion is considered to be a primary application of IVC. Here,
traffic instabilities will clearly lead to nonhomogeneous traffic
flows. This point can be clarified by looking at the
derivation. According to the convolution formula~\eqref{eq:convol},
the main result~\eqref{eq:prob_3} depends on the probability
distribution $P_2(\tau)$ for the time interval of the first
availability of a message, and the probability density $f_T$ for the
arrival time of the first destination vehicle.  According to
Eq.~\eqref{eq:prob_2}, $P_2(\tau)$ depends on the traffic situation in
the \textit{opposite} direction only, while $f_T$ as given by Eq.~\eqref{fT}
depends on the traffic flow in both directions, but only in the
\textit{target} region. Consequently, the analytical result remains
valid even in case of traffic congestion in the source region (or if
the message has been created and broadcasted by a standing vehicle),
as long as there are no traffic jams in the opposite direction
\textit{and} in the destination region. If traffic is congested in the
destination region, the system generally performs better than
predicted. Only if there is congested traffic in the \textit{opposite}
direction, the message propagation is significantly disturbed and the
analytical results are no longer applicable. However, this is not a
serious scenario as traffic information is important for the congested
lane rather than for the lane with freely flowing traffic.

\begin{figure}[!t]
\centering
\includegraphics[width=2.4in]{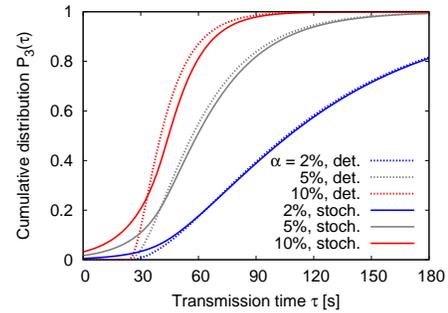}

 \caption{\label{fig:stoch}Transmission probability $P_3(\tau)$ as a
 function of the transmission time $\tau$ for a fixed direct
 communication range $r=\unit[200]{m}$ (thin lines), and for
 exponentially distributed ranges with the same expectation value
 $E(r)=\unit[200]{m}$, and 100\% correlation between the two hops.
 The values for $\rho_1$, $\rho_2$, $v_1$, $v_2$, and $r_{\text{min}}$
 are the same as in Fig.~\protect\ref{fig:trans_statistics}.  }
\end{figure}

In order to determine if the assumption of stochastic communication
ranges significantly changes the results compared to a fixed
communication range, we plot the corresponding
expressions~\eqref{eq:prob_3} and~\eqref{eq:prob_3dist} for
$\lambda_r=1/r$, i.e., assuming the same average communication range in
both models. Figure~\ref{fig:stoch} shows that for the practically
relevant regimes of low percentages $\alpha$ and high values of $P_3$,
the differences (of the order of one percent) are negligible. We will
therefore refer to the simpler fixed-range model~\eqref{eq:prob_3} in
the rest of this paper.  Note also that the stochastic model has a
higher connectivity for small values of $\tau$ but lower
connectivities later on with a crossover at about $P_3=0.25$. This is
to be expected since additional stochasticities typically smear out
distributions.

\begin{table}
\renewcommand{\arraystretch}{1.2}
 \caption{Characteristic quantities
 for the  efficiency of message dissemination via the opposite driving
 direction. The other  parameters are identical to the scenario illustrated in
 Fig.~\protect\ref{fig:trans_statistics}.}

\label{tab:trans_analyt}

\begin{tabular}{rccccccc}
\hline
$\alpha$  & $\langle \tau_2 \rangle$(s) & $\langle \tau_3 \rangle$(s) & $\tau_3^{(0.5)}$(s)& $\tau_3^{(0.9)}$(s) & $\tau_3^{(0.95)}$(s) & $v_p$(km/h) \\\hline
1\%  & 157  & 224  & 188  & 420  & 514 & 16.1 \\
2\%  & 90.7 & 124  & 106  & 222  & 269 & 29.0 \\
3\%  & 68.4 & 90.7 & 78.6 & 156  & 187 & 39.7 \\
5\%  & 50.7 & 64.0 & 56.7 & 103  & 122 & 56.3 \\
10\% & 37.3 & 44.0 & 40.4 & 63.6 & 73.0 & 81.8 \\
20\% & 30.7 & 34.0 & 32.2 & 43.8 & 48.5 & 105 \\
50\% & 26.7 & 28.0 & 27.3 & 31.9 & 33.8 & 128 \\
\hline
\end{tabular}
\end{table}

\section{\label{sec:sim_trans}Microscopic simulation results}
%
In this section, we compare the predictions of the analytic model of
Sec.~\ref{sec:model_trans} to simulation results obtained by a
microscopic traffic simulator that has been extended to support
inter-vehicle communication. Our principal interest is to check for
the reliability and robustness of our theoretical results based on the
assumption of exponentially distributed vehicles. How much error is
introduced by neglecting minimum safe distances between vehicles
and/or inhomogeneities of traffic flow, which clearly exist in real
traffic situations?

We have carried out a multi-lane traffic simulation of a~\unit[20]{km}
freeway stretch with two independent driving directions and one to
four lanes per direction.  The simulator uses the
\textit{Intelligent Driver Model}~\cite{Opus} as a simple, yet
realistic, car-following model, and the general-purpose lane-changing
algorithm MOBIL~\cite{MOBIL-TRR07}. A simplified demo version of this
simulator can be run interactively on the
web~\cite{Treiber-TrafficSimulator2010,traffic-simulation-de}. In order to introduce a minimal
degree of driver heterogeneity, the desired velocities of the
driver-vehicle units have been chosen Gaussian distributed with a
standard deviation of~\unit[18]{km/h} around a mean speed of
$v_0=\unit[120]{km/h}$ which is typical for a German freeway with
speed limits.  The other model parameters turned out not to be
relevant for the resulting statistics. Remarkably, this was even true
for the time headway parameter of the car-following model, although
this parameter directly influences the gaps between vehicles.
Furthermore, we used open boundary conditions with a constant inflow
$Q_{\text{in}}=\unit[1200]{/h/lane}$ at the upstream boundary. Note
that the traffic density is a result of the traffic dynamics. To
compare the simulation results with the analytical results, we
determined the traffic densities ($\rho_1$ and $\rho_2$) and average
speeds ($v_1$, $v_2$) for both driving directions by measurement via a
simulated loop detector.

For the purpose of this simulation study, we extended the traffic
simulator by a communication module.  After selecting the IVC vehicles
among all vehicles randomly according to the penetration level $\alpha$,
each IVC vehicle was equipped with a ``communication device'' which
allows for generating, sending, receiving, and storing
messages. Messages were generated on an event-driven basis that depended
on the considered IVC application (cf.\ Sec.~\ref{sec:appl_trans}
below). Here, test messages were generated whenever an IVC vehicle
passed a ``reference landmark'' at $x=\unit[10]{km}$. Furthermore, we
assumed a perfect communication channel: Messages were broadcasted
continually and vehicles in the opposite driving direction received the
messages instantaneously and without any transmission errors as soon
as they were within the communication range~$r$. Transversal distances
were neglected when determining whether a vehicle was within the
broadcast range. The message transport was simulated exactly as
described at the beginning of Sec.~\ref{sec:trans}.

\begin{figure*}[!th]
\centering
\subfloat{\includegraphics[width=2.2in]{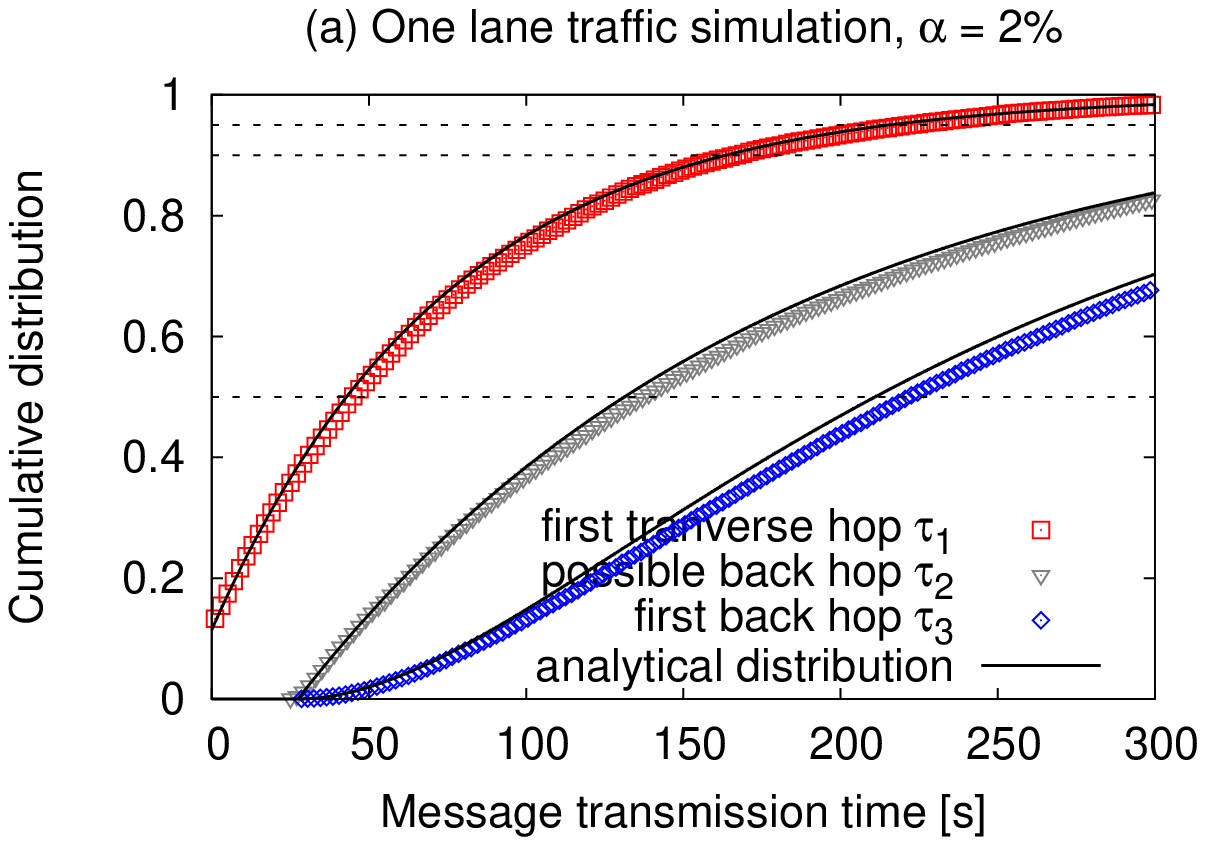}}
\subfloat{\includegraphics[width=2.2in]{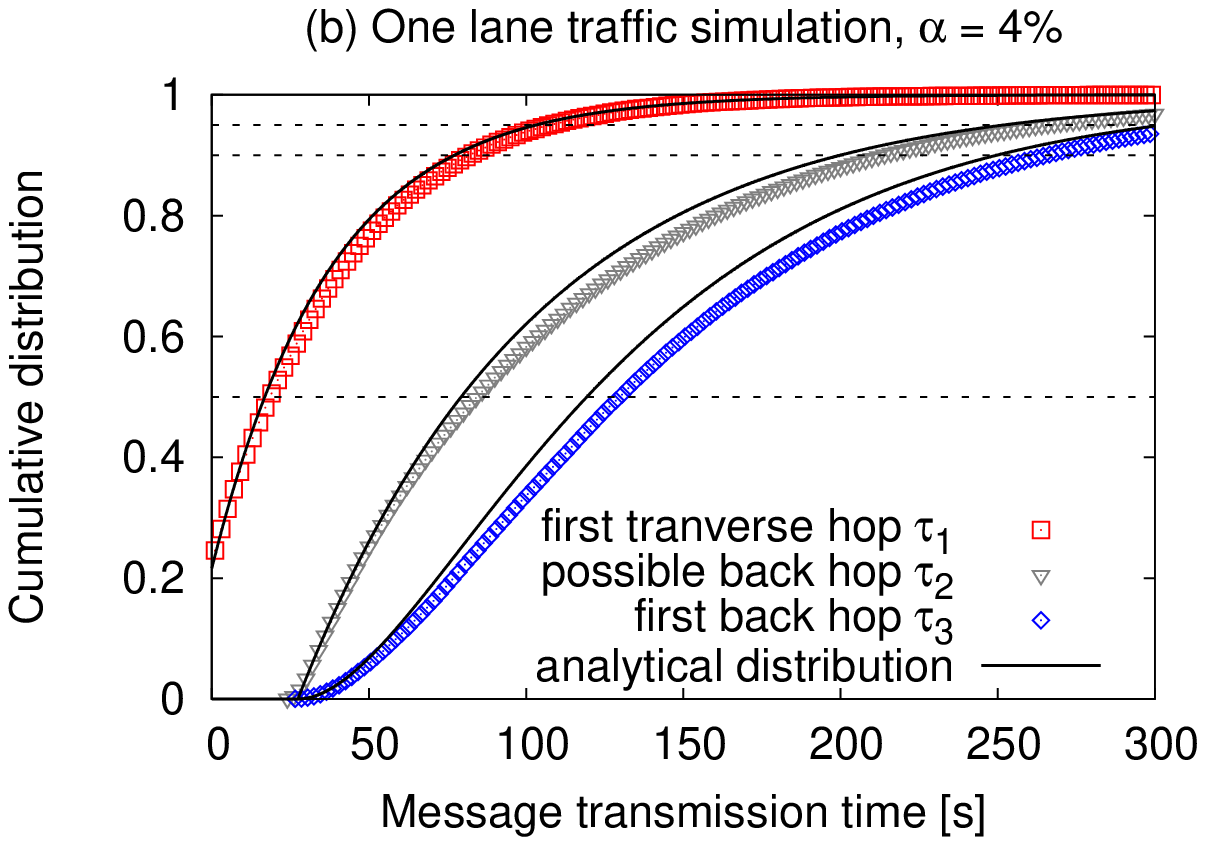}}\\[-0.1in]
\subfloat{\includegraphics[width=2.2in]{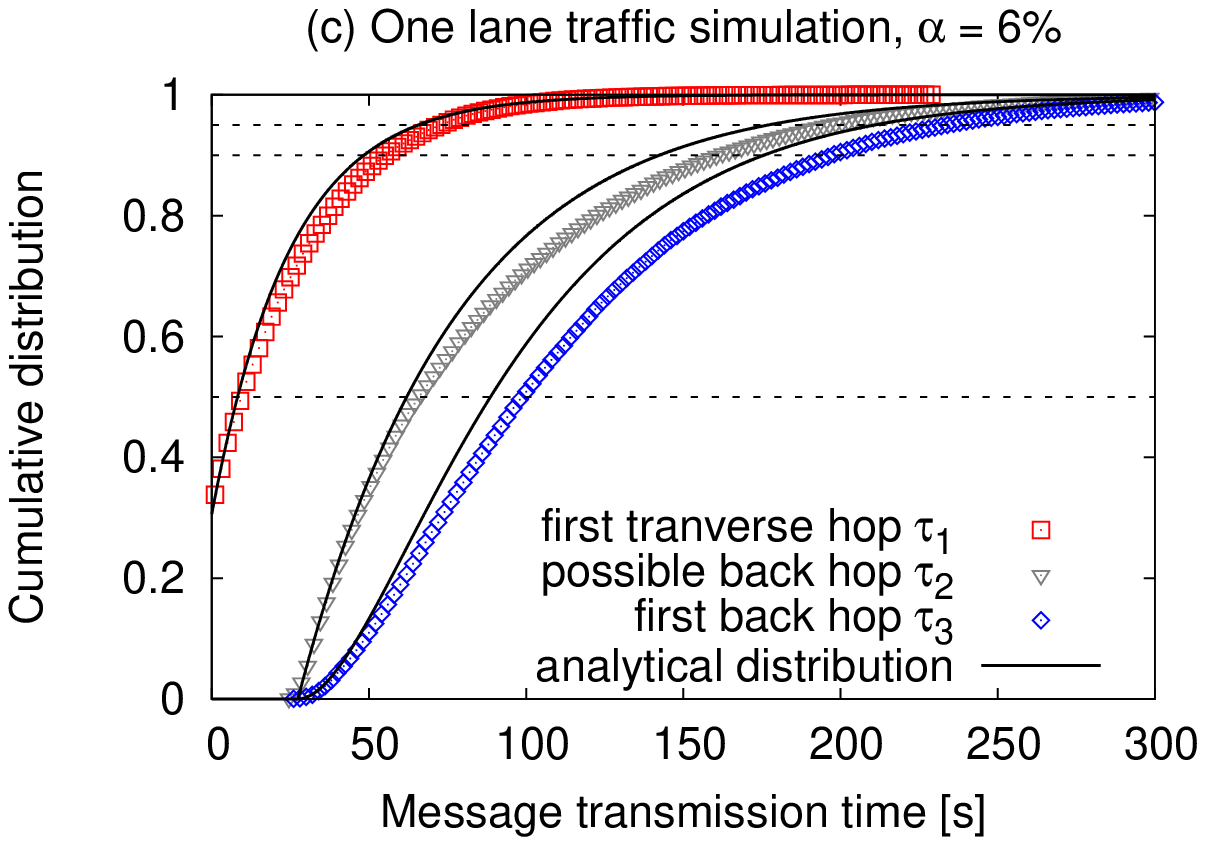}}
\subfloat{\includegraphics[width=2.2in]{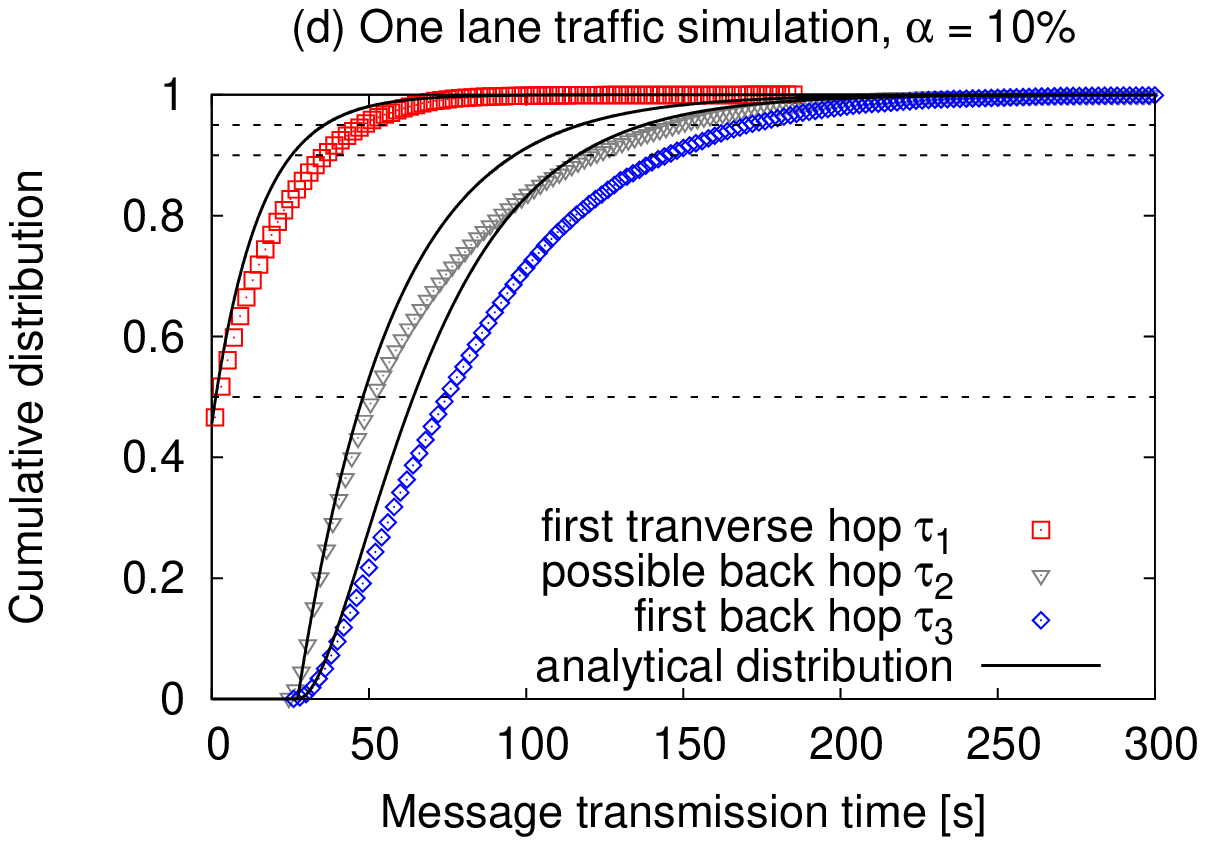}}

 \caption{\label{fig:trans_sim_1lane_alpha}Simulated vs. analytical
 results for the statistical properties of the characteristic
 communication times for four market penetrations on a road with one
 lane per direction (cf. Fig.~\protect\ref{fig:trans_screenshot}). The
 broadcast range $r$ is~\unit[200]{m}. The average speed $v$ (density
 $\rho$) of both driving directions was determined from the
 simulations approximately to~\unit[78.8]{km/h} (\unit[15.2]{/km}).}


\end{figure*}

Each simulation typically simulated several hours of real time and
generated about $100\,000$ message transmissions, serving as a solid
basis for the statistical
analysis. Figure~\ref{fig:trans_sim_1lane_alpha} shows the simulated
and the analytical distributions for the transmission times $\tau_1$,
$\tau_2$ and $\tau_3$ for one lane per direction and four different
penetration levels $\alpha$.  For low penetration levels, we have
observed a good agreement between the simulated and analytical results
while, for higher penetration levels of $\alpha=6\%$ and 10\%, the
median transmission times in the simulations were up to 10\% higher
than analytically predicted.

As a quantitative measure for the overall discrepancy, we introduce
the uncertainty measure
\begin{equation}
U = \frac{  \sum_i \left(x_i -\hat{x}_i \right)^2}{\sum_i \left(x_i -\bar{x}_i \right)^2}.
\end{equation}
Here, the sum runs over all message transmission times $i$ up to the
analytical 95\% quantile, $x_i$ and $\hat{x}_i$ denote the observed
and analytical values for the cumulative distribution of the relevant
time, respectively, and $\bar{x}$ is the arithmetic mean of the $x_i$
values. For the simulations shown in
Fig.~\ref{fig:trans_sim_1lane_alpha}, we have obtained the values
$U=0.006, 0.013, 0.030$ and 0.084 for $\alpha=2\%, 4\%, 6\%$, and
10\%, respectively. This finding confirms that the agreement is best
for low penetration levels.

\begin{figure}[!t]
\centering
\includegraphics[width=\linewidth]{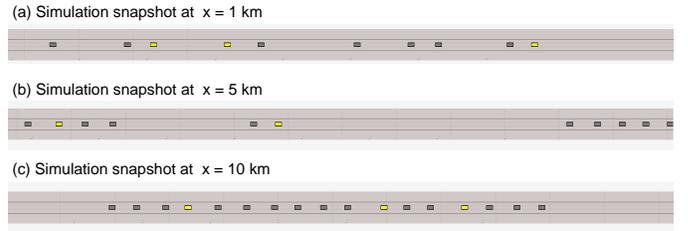}

 \caption{\label{fig:trans_screenshot}Screenshots of one driving
 direction in a single-lane traffic simulation at three locations
 showing the clustering of vehicles in platoons due to distributed
 desired velocities in combination with the lack of overtaking
 possibilities. For the purpose of illustration, the market
 penetration of equipped (labeled as yellow) vehicles was set to
 20\%.}

\end{figure}

When looking at the actual vehicle positions observed in the
simulations, the reason for the deviations at higher penetration
levels becomes obvious. Figure~\ref{fig:trans_screenshot} shows
snapshots of the simulator for one driving direction with a single
lane. Due to the distributed desired velocities and the lack of
overtaking possibilities, vehicles that initially enter the road with
identical headways tend to cluster behind slower vehicles. At
$x=\unit[10]{km}$, i.e., at the point of message generation, this
results in significant vehicle platoons. Both the cluster-forming
process and the minimum safety gap between the vehicles make the
arrival process significantly \textit{non-Poissonian}. In particular,
the large gaps between the clusters lead to larger transmission times
in the simulations as compared to our previous analytical
results. Nevertheless, for sufficiently low penetration levels, the
average distance between equipped vehicles becomes comparable to the
typical cluster size. As a consequence, the Poissonian assumption can
be satisfied for the
\textit{IVC equipped vehicles}, even if this is not the case for \textit{all}
vehicles. This explains the good agreement of the theoretical
prediction for \textit{low} penetration levels and shows that the
analytical model of Sec.~\ref{sec:model_trans} is quite robust with
respect to violations of its assumptions.

\begin{figure*}[!t]
\centering
\subfloat{\includegraphics[width=2.2in]{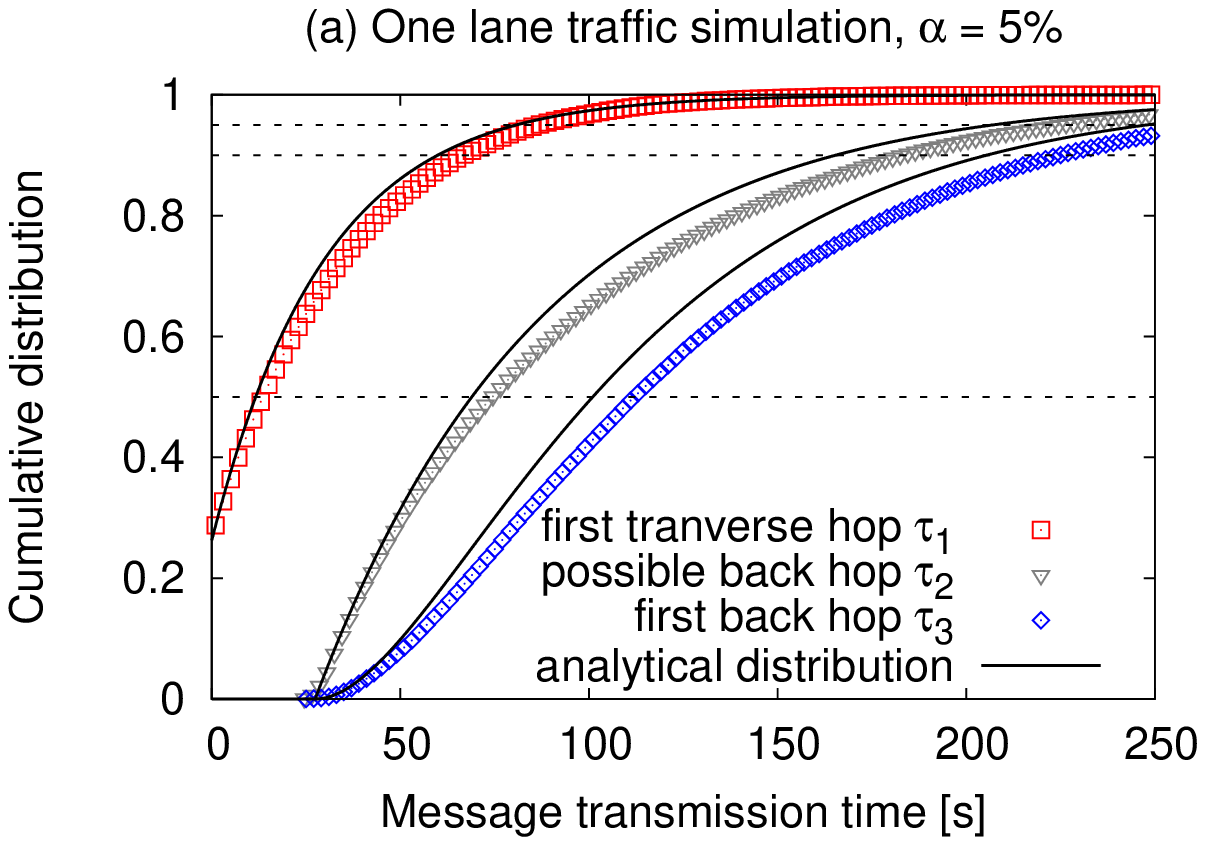}}
\subfloat{\includegraphics[width=2.2in]{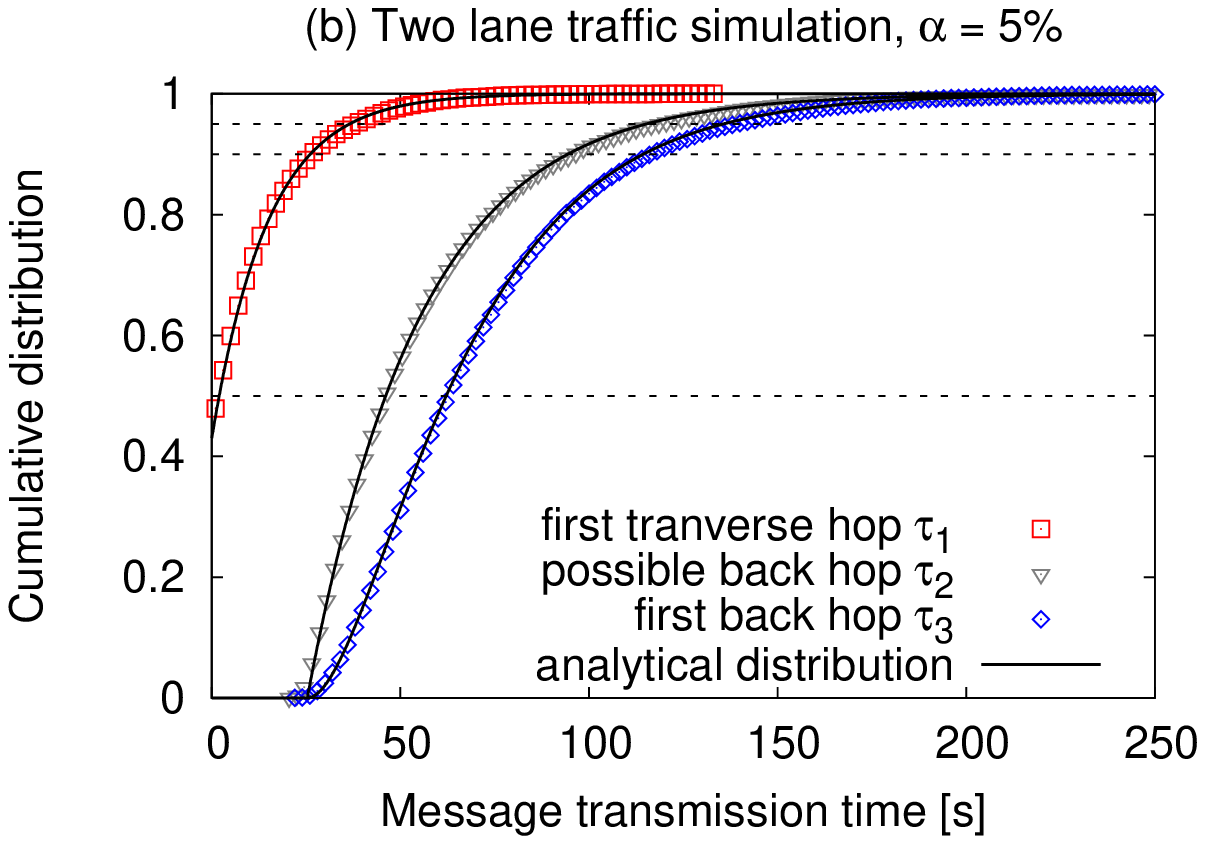}}\\[-0.1in]
\subfloat{\includegraphics[width=2.2in]{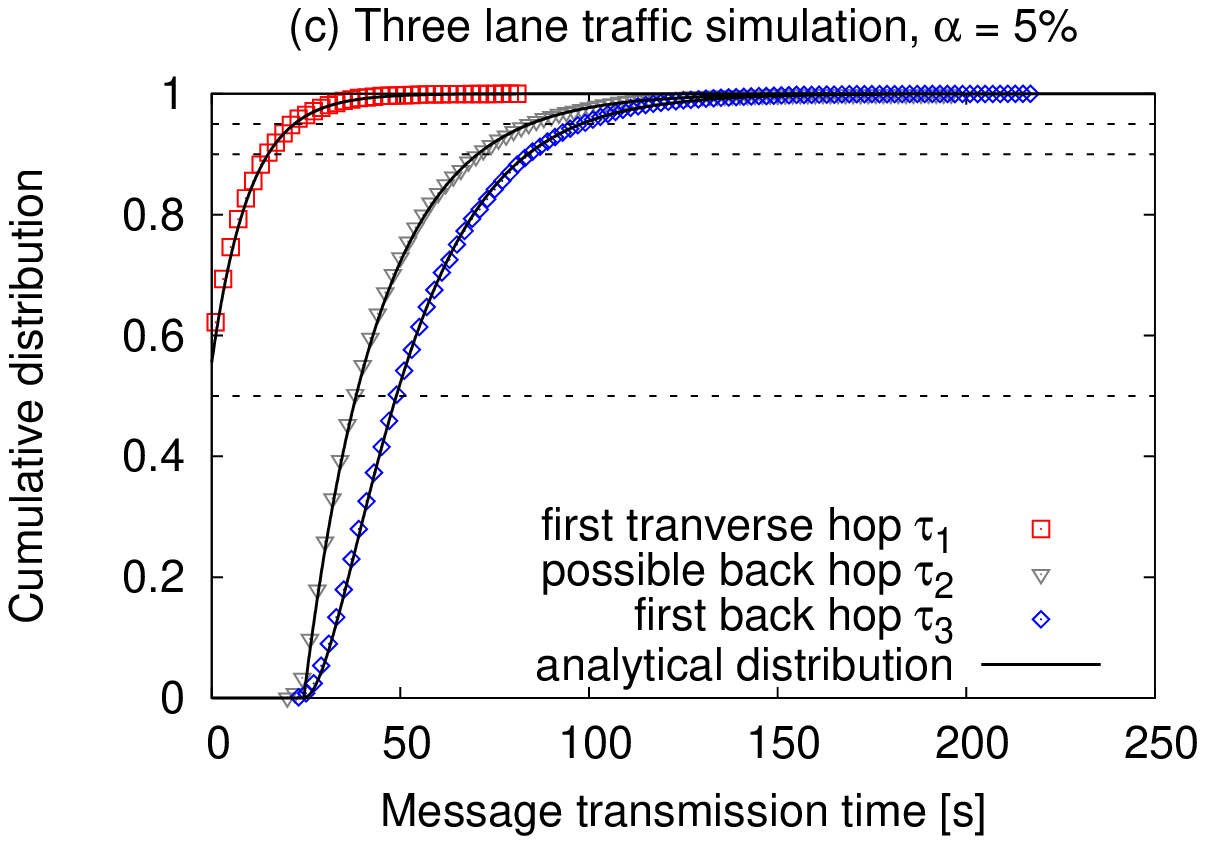}}
\subfloat{\includegraphics[width=2.2in]{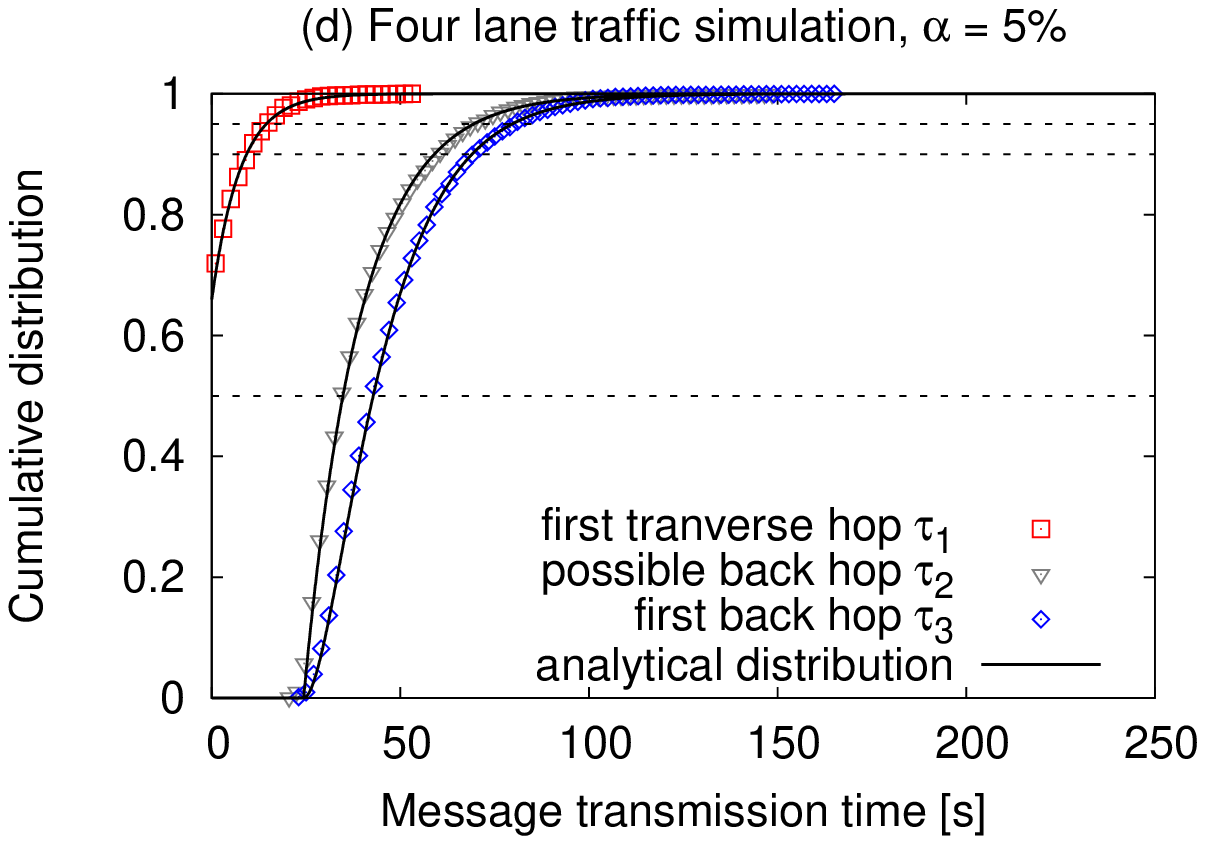}}

 \caption{\label{fig:trans_sim_lanes}Simulated and analytical results
 for the statistical properties of the characteristic communication
 times, studying roads with one to four lanes per direction. The market
 penetration $\alpha=5\%$ and the broadcast range $r=\unit[200]{m}$
 are kept constant while the average speed and density of both driving
 directions were determined from the simulations
 ($v=\unit[78.8]{km/h}$, $\rho=\unit[15.2]{/km}$ for 1 lane;
 $v=\unit[85.4]{km/h}$, $\rho=\unit[28.1]{/km}$ for 2 lanes;
 $v=\unit[88.7]{km/h}$, $\rho=\unit[40.6]{/km}$ for 3 lanes;
 $v=\unit[88.8]{km/h}$, $\rho=\unit[53.6]{/km}$ for 4 lanes).}


\end{figure*}

Finally, Figure~\ref{fig:trans_sim_lanes} shows a comparison between
the analytical model and simulated message propagation for different
numbers of lanes per direction, while keeping the penetration level at
a constant value $\alpha=5\%$. It turns out that the discrepancies
vanish almost completely, when there is more than one lane. While the
uncertainty measure was $U=0.024$ for one lane, we observed $U<0.0004$
for two to four lanes. In terms of cumulative distribution functions,
the error is less than $\sqrt{U}=2\%$.  The same applies to the
relative error of the median (or other quantiles) of the message
transport times. The primary reasons for this good agreement are the
overtaking opportunities preventing the build-up of large vehicle
platoons, at least, if no traffic breakdowns occur.  Generally, we
found that the error decreases with an increasing number of lanes and
with a decreasing penetration level.

\section{\label{sec:appl_trans}IVC for jam front detection and traveler information}
\begin{figure*}[!t]
\centering
\includegraphics[width=4in]{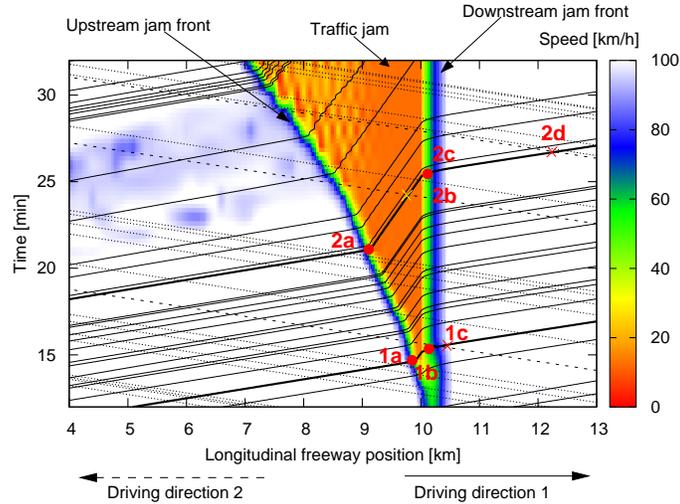}
 
  \caption{\label{fig:jam_fronts}Space-time diagram of a scenario with
  traffic congestion. The contour plot shows the average speed of all
  vehicles in driving direction~1. Trajectories of equipped vehicles
  (only 1\% of all vehicles) are displayed by solid and dashed lines,
  depending on the driving direction. Points indicate the generation of
  messages when detecting the shock fronts of the jam and crosses
  indicate message transmissions.}

\end{figure*}

In this section we demonstrate the efficiency of the transverse
hopping mechanism by simulating the propagation of messages regarding
traffic conditions that are created by vehicles entering and/or
leaving a traffic jam. The fronts of the jam can be autonomously
detected by individual vehicles based on the decrease or increase in the average driving
speed~\cite{Arne-ACC-TRC,ChakravarthyPileUpCrash_IVC}. We have simulated a bi-directional
roadway with two lanes in each driving direction and with a
bottleneck at $x=\unit[10]{km}$ in one driving direction. While traffic is free in one
direction, traffic becomes congested in the other due to an increasing demand that
exceeds the capacity at the bottleneck (peak-hour scenario with time-varying boundary conditions). The resulting spatiotemporal
contour plot of the average speed is shown in
Fig.~\ref{fig:jam_fronts}. Furthermore, the trajectories of the
IVC-equipped vehicles are displayed by solid and dotted lines
depending on the driving direction. Notice that the market penetration
is only~1\%.

In our simulation, the breakdown of traffic flow occurs approximately
after~\unit[14]{min} at the bottleneck location. For the purpose of
illustration, we discuss the involved processes by means of two
equipped vehicles which are marked in Fig.~\ref{fig:jam_fronts} by
thicker lines labeled as ``1'' and ``2''. First, the upstream end of
the jam is detected by the decrease in speed (corresponding to the
change in the slope of the trajectory) which is labeled as ``1a'' in
the space-time diagram. Second, the same vehicle records the
downstream end of the traffic jam while accelerating to its desired
speed, thereby creating the message ``1b''. The same processes apply
to car 2 labeled as ``2a'' and ``2c''. These messages are received by
equipped vehicles in the opposite driving direction within the
communication range, denoted as events ``1c'', ``2b'' and ``2d'',
respectively. Finally, the messages are transported in upstream
direction and broadcasted, until they are received by vehicles in
lane~1 at later times. In Fig.~\ref{fig:jam_fronts}, this corresponds
to the intersection of trajectories of relay vehicles (thicker
points) with a solid-line trajectory.

In order to be useful for traveler information services, the messages
have to propagate faster than the fronts of the traffic jam. It is
known from empirical observations that the head of a traffic jam (the
``downstream jam front'') is either fixed at the bottleneck or moves
upstream with a characteristic speed
of~\unit[15$\pm$5]{km/h}. Moreover, possible propagation velocities
for the jam front moving in upstream direction range from~0 to
about~\unit[18]{km/h}, depending on the current
inflow~\cite{Helb-Opus}. Note that in case of dissolving congestion,
the upstream jam front can also propagate in \textit{downstream}
direction. In the simulation example shown in
Fig.~\ref{fig:jam_fronts}, the upstream front of the traffic jam moves
in upstream direction with a speed of about~\unit[10]{km/h}, while the
downstream front is fixed at the (stationary) bottleneck.

In Eq.~\eqref{eq:average_msg_speeds} of Sec.~\ref{sec:model_trans}, we
have defined the average speed of information dissemination $v_p$ with
respect to the first vehicle that received this message in the target
region located upstream of the source by at least a distance
$r_\text{min}$. Nevertheless, the same message is also useful for
later vehicles in the destination region, at least for some time. Since,
during this time, the message is transported by the relay vehicle with
a speed larger that $v_p$, Eq.~\eqref{eq:average_msg_speeds} denotes
the \textit{worst case}. For example, the messages ``1a'' and ``1b''
referring to the situation at $t =
\unit[15]{min}$ are received by several vehicles within the
following~2 to \unit[3]{minutes}. From the end user's point of
view, a ``usability measure'' could consider the up-to-dateness of the
information about the upstream and downstream jam fronts at a given
distance from the jam, e.g.,  \unit[1]{km}. At this distance,
the messages 1a and 1b are the
most recent ones for three vehicles (entering the section shown in
Fig.~\ref{fig:jam_fronts} at times between \unit[15]{min} and
\unit[16.5]{min}). For these vehicles, the ``age'' of the upstream
information 1a lies between \unit[2.3]{min} and \unit[3.1]{min}.  The downstream information 1b is even more recent. For the set of 9~vehicles located between message 1a and 2a, the average message age is \unit[2.5]{min}. 

Note that one can also take into account that a traffic management
center or the police may send out ``public'' messages about the
current traffic state, roadwork or incident conditions, complementing
messages created by autonomous vehicles. Representing this by a
standing vehicle in the shoulder lane for the initial message
broadcast, this will not change the further transmission times. A
standing vehicle would correspond to a horizontal trajectory in
Fig.~\ref{fig:jam_fronts}.

Finally, we point out that messages regarding jam front
positions are up-to-date as long as the \textit{collective} traffic
dynamics does not change significantly. While the downstream jam front
is fixed at the bottleneck (and therefore easy to predict at later
times), the moving upstream front can be estimated with an accuracy of
several~\unit[100]{m}. However, data-fusion of \textit{several}
messages and model-based prediction can be used to reduce these errors
dramatically~\cite{Kesting-StateEstimation-TGF09}. For a quantitative analysis of such a vehicle-based
jam-front detection, we refer to Refs.~\cite{Kesting-StateEstimation-TGF09,IVC-Martin-TRR07}.

\section{\label{sec:diss}Discussion and Conclusions}
Inter-vehicle communication (IVC) based on vehicular ad hoc networks
(VANETs) is a promising and scalable concept for exchanging
traffic-related information among vehicles over relatively short
distances. Advanced traveler information systems are recognized as an
important application of this decentralized approach in the first
deployment phase, as IVC will provide local and up-to-date traffic
information. Apart from the drivers appreciating such reliable and
up-to-date traffic information, future driver assistance and safety
systems may benefit from IVC as well, for example, by issuing warnings
when a traffic jam or an accident is several hundred meters ahead.

Adaptive cruise control can automate the braking and acceleration of a
car. Processing of non-local information received via IVC could help
future ``traffic-adaptive'' cruise control systems to anticipate the
traffic situation and therefore to automatically adapt the driving
style to it, increasing driving comfort, safety and traffic
performance~\cite{Arne-ACC-TRC,Arem_ACC_Impact_Transaction}. Notice that the sensor technology
needed for driver assistance systems can be used to automatically
produce messages on an event-oriented basis.

However, like all technologies relying on local communication, IVC
faces the ``penetration threshold problem''. Thus, the system is
effective only if there is a sufficient number of communication
partners to propagate the message between equipped cars. Therefore, it
is crucial to assess the feasibility of different communication
variants in terms of the necessary critical market penetration. 

In this paper, we have investigated this aspect both, analytically and
by simulation for a basic strategy of message dissemination by
``transverse hopping''. In this mode, equipped vehicles in the
opposite driving direction are used to transport the messages serving
as relays. Our results indicate that the transverse hopping mechanism
is favorable in the first stages of the deployment of an cooperative
IVC system, since it is already effective for market penetration
levels as low as 1-2\%. 

In order to gain more insights into the factors influencing the
reliability and effectiveness of the ``store-and-forward'' message
propagation in IVC systems, we derived analytical expressions for
communication delay times. An important result of this paper is that
the analytically derived statistics have been confirmed by extensive
traffic simulations. Although the main assumptions made in deriving
these models -- homogeneous traffic and exponentially distributed
inter-vehicle distances -- are normally not perfectly met in real
traffic flow, it turns out that the theoretical results are remarkably
robust with respect to violations of the models' assumptions. In
particular, for multi-lane traffic and penetration levels below~5\%,
the errors are typically a few percent only. Notice that, in turn, the
analytical expressions may also serve to test or validate new
implementations of communication modules in traffic simulation
software. 

One may wonder why the idealized assumptions work particularly
well in multi-lane traffic which clearly is more complex than
single-lane traffic. This is mainly caused by the possibility to
change lanes and
overtake thereby dissolving platoons behind slow vehicles. The positional
correlations implied by platoons, in turn,
are the main cause why the assumption of the Poissonian positional
distribution is not satisfied. Other kinds of complexity, however,
lower the reliability of this assumption. For example, in stop-and-go
traffic or at intersections the overall traffic flow varies strongly
(typically on scales of \unit[1]{min} per \unit[1]{km}). Such
non-stationary traffic conditions can only be approximately treated as
a superposition of the prevailing traffic densities.

Since our focus was on traffic dynamics rather than on the details of
the communication protocols, we generally assumed idealized
communication conditions, i.e., instantaneous and error-free
transmission below a certain direct communication range and a failure
rate of 100\% above.  However, we have shown analytically that the
transmission properties change in a predictable way when assuming more realistic
communication conditions. For example, relaxing the assumption
of a fixed communication range by treating the communication range as
a random variable resulted in little changes (Fig.~\ref{fig:stoch}).
Furthermore, assuming a complete failure of the communication
chain with a  probability $f$ simply will
 reduce the market penetration parameter $\alpha$ by a factor $1-f$,
at least,  as long as the failures can be considered to be uncorrelated.

The assumption that a vehicle continuously broadcast messages could be
relaxed by considering a periodic broadcast of messages in time
intervals $\tau_t$. This would result in two additional convolutions of the relevant
transmission time $\tau_3$ with the uniform distribution in
the interval $[0,\tau_t]$. This essentially increases the median of
the communication time by  $2\tau_t/2=\tau_t$.

It remains to be shown to which extent other imperfections of real communications (such as delay times
for establishing a direct communication, failures due to high
relative velocities , or channel conflicts and contention for an
increasing penetration level) will influence the
results. In any case, the microscopic details of DSRC communication require network simulation software such as ns2.

For demonstration purposes, we considered a congestion-warning
application in a complex traffic simulation which is operative for
penetration rate as low as 1\%. However, IVC is only one building
block of a future integrated traffic communication system. As a
straightforward next step, including police cars and emergency
vehicles into the IVC fleet will lead to a timely production of
event-related messages. Furthermore, adding infrastructure-vehicle
communication to the system may help to overcome the penetration
barrier~\cite{Ma-C2I-2009}. This will be particularly economic and efficient when placing
the infrastructural communication units near to known bottlenecks,
where the necessary sensors for producing event-based traffic messages
are already in place.
\section*{Acknowledgments} 
A.K. kindly acknowledges financial support from the Volkswagen~AG
within the German research initiative AKTIV.
 D.H. has been partially supported by the ``Cooperative Center for Communication Networks Data
Analysis'', a NAP project sponsored by the Hungarian National Office
of Research and Technology under grant No.\ KCKHA005.


\end{document}